# Fermi surface reconstruction and multiple quantum phase transitions in the antiferromagnet CeRhIn$_5$

Lin Jiao[a], Ye Chen[a], Yoshimitsu Kohama[b], David Graf[c], E. D. Bauer[b], John Singleton[b], Jian-Xin Zhu[b], Zongfa Weng[a], Guiming Pang[a], Tian Shang[a], Jinglei Zhang[a], Han-Oh Lee[a], Tuson Park[d], Marcelo Jaime[b], J. D. Thompson[b], Frank Steglich[e,a], Qimiao Si[f,1], H.Q. Yuan[a,g,1]

[a]*Center for Correlated Matter and Department of Physics, Zhejiang University, Hangzhou, Zhejiang 310058, China*

[b]*Los Alamos National Laboratory, Los Alamos, NM 87545, USA*

[c]*National High Magnetic Field Laboratory, Florida State University, Tallahassee, Florida 32310, USA*

[d]*Department of Physics, Sungkyunkwan University, Suwon 440-746, Korea*

[e]*Max Planck Institute for Chemical Physics of Solids, 01187 Dresden, Germany*

[f]*Department of Physics and Astronomy, Rice University, Houston, TX 77005, USA*

[g]*Collaborative Innovation Center of Advanced Microstructures, Nanjing University, Nanjing 210093, China*

[1] To whom correspondence should be addressed: hqyuan@zju.edu.cn or qmsi@rice.edu

*Abstract*

**Conventional, thermally-driven continuous phase transitions are described by universal critical behaviour that is independent of the specific microscopic details of a material. However, many current studies focus on materials that exhibit quantum-driven continuous phase transitions (quantum critical points, or QCPs) at absolute zero temperature. The classification of such QCPs and the question of whether they show universal behaviour remain open issues. Here we report measurements of heat capacity and de Haas-van Alphen (dHvA) oscillations at low temperatures across a field-induced antiferromagnetic QCP ($B_{c0} \approx 50$ T) in the heavy-fermion metal CeRhIn$_5$. A sharp, magnetic-field-induced change in Fermi surface is detected both in the dHvA effect and Hall resistivity at $B_0^* \approx 30$ T, well**

**inside the antiferromagnetic phase. Comparisons with band-structure calculations and properties of isostructural CeCoIn$_5$ suggest that the Fermi-surface change at $B_0^*$ is associated with a localized to itinerant transition of the Ce-4$f$ electrons in CeRhIn$_5$. Taken in conjunction with pressure experiments, our results demonstrate that at least two distinct classes of QCP are observable in CeRhIn$_5$, a significant step towards the derivation of a universal phase diagram for QCPs.**

Intermetallic heavy-fermion metals, the ground states of which can be tuned readily by control parameters other than temperature, such as magnetic field, pressure or chemical composition, form attractive systems for the study of QCPs (1, 2). Despite examples of QCPs in many heavy-fermion materials, their theoretical classification, analogous to that for thermally-driven phase transitions, is yet to be achieved (2, 3). An extension of the theory of thermally-driven critical points to the zero-temperature limit predicts that only fluctuations of an order parameter (for example, the sublattice magnetization of a spin-density wave (SDW)) are singular in space and time at a critical value of the control parameter (4-6). The Fermi surface evolves smoothly across the QCP. Though there is some experimental support for this model of a QCP, e.g., in CeCu$_2$Si$_2$ (7), the quantum critical response of some other heavy-fermion compounds is clearly inconsistent with its predictions (8-10). Going beyond this conventional Landau framework, a qualitatively different model predicts a sharp reconstruction of the Fermi surface while crossing the QCP due to the essential involvement of the electronic degrees of freedom (11-13). Such an *unconventional* QCP has been proposed to involve the critical destruction of the Kondo effect, in addition to fluctuations of the order parameter (3). Behaviour that supports this type of QCP has been found in various heavy-fermion systems, including $UCu_{5-x}Pd_x$ (8), $CeCu_{6-x}Au_x$ (9), and $YbRh_2Si_2$ (10). In each material, however, direct evidence for a change in Fermi surface is lacking. Besides the needs to verify the basic predictions of this model of QCPs, a further test of its validity is the more restrictive observation of a change in the Fermi surface as a function of multiple tuning parameters.

$CeRhIn_5$, a heavy-fermion antiferromagnet with a Néel temperature $T_N \approx 3.8K$ at ambient pressure (14), is very suitable for this purpose. As shown in the schematic phase diagram at zero temperature (Fig. 1A), application of pressure suppresses the antiferromagnetic (AF) order and induces superconductivity (15, 16). In the presence of a modest magnetic field, sufficient to suppress superconductivity, an AF QCP is exposed through pressure tuning (15, 16). A distinct change of the Fermi surface is observed across this pressure-induced QCP via dHvA oscillations (Fig. 1A, blue arrow) that find a large Fermi surface, like that of $CeCoIn_5$, above the QCP (17). In addition, the AF order of $CeRhIn_5$ is robust against magnetic field at ambient pressure (18). Thus, $CeRhIn_5$ permits measurements of magnetic quantum oscillations across its field-tuned QCP at ambient pressure, providing a rare system in which the nature of QCPs can be probed using more than one tuning parameter.

Here, we report our study of the phase transitions in $CeRhIn_5$ at ambient pressure using isothermal measurements of the *ac* heat capacity and dHvA oscillations in pulsed magnetic fields of up to 72 T (see Fig. 1B), plus Hall resistivity and dHvA oscillations in a *dc* field of up to 45 T [see Supporting Information (SI) Appendix]. These experiments allow explicit mapping of the magnetic field-temperature phase diagram and the investigation of the evolution of the Fermi surface as a function of magnetic field.

4## Results

### Determination of the magnetic field-temperature phase diagram

Figure 2A displays the three-dimensional (3D) plots of the heat capacity coefficient $C_p/T$ as a function of magnetic field and temperature for CeRhIn$_5$ with $B//c$. Similar results were also observed for $B//a$ (see SI Appendix, Fig. S3). The pronounced maximum in $C_p(B)/T$ marks the onset position $B_c(T)$ of the suppression of the Néel transition by the magnetic field. The magnetic phase boundary $T_N(B)$ of CeRhIn$_5$ is then derived by projecting $B_c(T)$ onto the $B$-$T$ plane; the corresponding phase diagrams ($B//a$ and $B//c$) are shown in Fig. 2B. At low fields, the boundaries are consistent with heat-capacity results in fields provided by a Physical Properties Measurement System (PPMS, Quantum Design). These heat-capacity data, along with the field-dependence of the magnetic susceptibility ($B//a$) (Fig. 1B and also SI Appendix, Fig. S4) yield a $T = 0$ critical field $B_{c0}$ [$\equiv B_c(T \rightarrow 0)$] of about 50 T. In the low temperature limit, the heat-capacity coefficient $C_p(B)/T$ is substantially enhanced in the paramagnetic phase ($B > B_c$) in comparison with that well inside the AF state (see the inset of Fig. 2B).

### Evidence for a field-induced reconstruction of Fermi surface

Well inside the AF state, a field-induced sharp change of Fermi surface is evidenced in the Hall resistivity $\rho_{xy}(B)$ and the dHvA oscillations. Figure 3A displays $\rho_{xy}(B)$ at various temperatures. Here the field was applied along the $c$-axis; because of the flat sample geometry, it was difficult to measure the Hall resistivity for fields applied within the $ab$-plane. The jump in $\rho_{xy}(B)$ around $B_M \approx 18$ T is attributed to a metamagnetic transition with the sample alignment slightly away from the $c$-direction; $B_M$ increases as a function of $1/\cos\theta$ when the field is tilted by $\theta$ from [100] to [001], as found in other measurements (18). Upon further increasing the magnetic field, the Hall resistivity $\rho_{xy}(B)$ undergoes a sharp change at $B^* \approx 31\pm1$ T, which disappears at temperatures above 1.4 K. Around $B = B^*$, $\rho_{xy}(B)$ changes slope significantly, suggesting a pronounced jump in the differential Hall coefficient $R_H = d\rho_{xy}/dB|_{B=B^*}$. For 20 T$<B<$30 T, $\rho_{xy}(B)$ decreases linearly with increasing field, indicating electron-type majority carriers. For $B>33$ T, the negative slope of $\rho_{xy}(B)$ is substantially reduced compared to that at $B<30$ T, and there is a significant change in the curvature at higher fields, typical behavior of a multiband conductor. The jump of Hall coefficient at $B^*$ suggests a sharp change of Fermi surface.

In Fig. 3B, we show Fourier transforms of the dHvA oscillations obtained at $T = 0.33$ K for $B//c$, which were measured by using a torque technique (CuBe cantilever) in $dc$ fields of up to 45 T. For $B<30$ T, several dHvA frequencies and their harmonics are well identified, which coincide with previous results (19). With increasing field, new (or shifted) dHvA frequencies, labeled $\alpha_1$, $\alpha_2$ and $\beta_1$, show up above $B^* \approx 30$ T. Figure 3C plots the field evolution of the dHvA amplitudes of the $\alpha_1$ branch at several temperatures; the $\alpha_2$ and $\beta_1$ branches show similar behavior. One can see that the dHvA amplitudes suddenly vanish around $B^* \approx 30$ T at the lowest temperatures, providing a precise determination of the onset field $B^*$ for the Fermi surface reconstruction. The dotted line shows a fit of the data to the Lifshitz-Kosevich (LK) formula (20) for fields above 30 T. The LK formula predicts that the high-frequency oscillations should continue to be observable (i.e. have an amplitude well above the noise floor) to fields well below 30 T, which is inconsistent with our experimental data. With increasing temperature, the dHvA amplitude is reduced and the critical field $B^*$ slightly shifts to higher temperature. Above $T \approx 1.4$ K, where the Hall jump disappears, the new dHvA oscillations also cannot be resolved. Thus, both the Hall data and the dHvA oscillations consistently provide strong evidence of a Fermi surface reconstruction around $B^* \approx 30$ T at our base temperatures, with no indication of reconstruction with increasing temperature above $T \approx 1.4$ K.

Evidence for an isotropic reconstruction of Fermi surface can be inferred from measurements with fields applied along the *a*-axis. Figure 4A shows the results of dHvA oscillations measured with a piezo-cantilever in quasistatic fields provided by a 45 T hybrid magnet (*B*//*a*). This measurement is extremely sensitive to tiny changes of the magnetization of the sample and thus provides a largely enhanced resolution. Unfortunately, due to experimental constraints, we could only measure the sample up to 38 T at one temperature ($T = 0.31$ K). After subtracting the background, a pronounced change of dHvA oscillations is seen clearly around $B^* \approx 30.5$ T (upper inset of Fig. 4A). In the lower inset of Fig. 4A, we plot the dHvA signals as a function of $1/B$ near the maximum field, which demonstrate periodic oscillations with a rather large frequency of around 10600 T. As an alternative depiction, the main panel plots the amplitude of the high-frequency dHvA oscillations averaged over 2 T windows. Resembling that of *B*//*c*, the collapse of dHvA amplitude at $B^* \approx 30.5$ T marks an abrupt change of Fermi surface; the so-derived value of $B^*$ is very close to that for *B*//*c*, demonstrating a rather isotropic value of $B^*$ for both *B*//*a* and *B*//*c*.

In order to further examine the field-induced reconstruction of Fermi surface, we used an induction method to extend the measurements of dHvA oscillations in a 75 T pulsed magnet (*B*//a). These higher fields provide a much broader window for the analysis of Fourier transforms than that of the *dc* field data. Figure 4B shows Fourier transforms of the dHvA oscillations obtained at $T = 0.5$ K for field windows of 10 T <*B*

< 30 T ($B<B^*$) and 45 T<$B$<70 T ($B>B^*$), respectively. The peaks, labeled $f_{a1}$-$f_{a4}$, in the dHvA spectrum are located within the frequency range 200 T-1000 T for fields of 10 T <$B$ < 30 T. They coincide well with our preceding data obtained in the 45 T hybrid magnet as well as previous measurements (19), and are characteristic of a 'small' Fermi surface that does not include the 4$f$-electrons of Ce. Several higher harmonics of the most pronounced dHvA frequencies ($f_{a4} \approx$ 790 T) are observed. Four new dHvA branches ($f_{a5}$-$f_{a8}$) with much larger frequencies (7 kT<$f$<15 kT) occur at $B>B^*$; the frequency of $f_{a7}$ is close to that obtained in a $dc$ field. These new dHvA frequencies persist up to $T$ = 0.8 K but become difficult to resolve around $T$ = 1 K. It is noted that, owing to their short duration, pulsed experiments may give lower signal-to-noise ratios than equivalent dc field measurements. Hence, lower temperatures are needed to resolve all the dHvA oscillations.

**Summary and discussion**

Figure 5A presents the temperature–magnetic field phase diagram of CeRhIn$_5$ for $B//a$ and $B//c$ derived from our experimental data. On applying a magnetic field, the AF transition temperature $T_N$ is suppressed continuously to the lowest accessible temperature ($T \approx$ 0.4 K for $B//a$), providing strong evidence for an AF QCP around $B_{c0} \approx$ 50 T. The field dependence of $T_N$, i.e., $T_N \sim (B_{c0}-B)^{2/3}$, suggests an SDW-type QCP. Inside the AF phase, the electronic system undergoes a sharp reconstruction of Fermi surface at $B^*(T)$, changing from a local-moment antiferromagnetic order (AF$_S$) to an SDW of the large Fermi surface (AF$_L$). Further experiments are needed to determine whether the $B^*$ line eventually fades away in the paramagnetic state or terminates at a critical end point.

High-frequency dHvA oscillations potentially could emerge with increasing magnetic field due either to the so-called Dingle damping factor or to magnetic breakdown (20). However, it is clear from Fig. 3 and Fig. 4A that the higher frequencies appear *suddenly* at $B^*$, rather than growing smoothly. If the Dingle factor were responsible for their growth, each oscillation should gradually become visible over differing ranges of field. The scenario of magnetic breakdown at $B^*$ also can be excluded because the new dHvA frequencies appear at indistinguishable fields for $B//a$ and $B//c$; in an anisotropic band structure, such as that of $CeRhIn_5$, any breakdown fields also would be anisotropic. Thus, all these suggest that the sudden emergence of new dHvA frequencies at $B^*$ corresponds to a field-induced reconstruction of Fermi surface in $CeRhIn_5$.

Modifications of the magnetic structure in a magnetic field, e.g., a metamagnetic transition, may induce a reconstruction of Fermi surface. With the current resolution of our experiments, we cannot detect any thermodynamic phase transition at $B = B^*$ either in magnetic susceptibility or heat capacity (cf. Fig. 1B). Furthermore, the critical field of a metamagnetic transition usually varies with the field orientation as shown for the transition at $B_M$. Our observations of an isotropic value of $B^*$ for $B//a$ and $B//c$ disfavor such a scenario. Therefore, it is unlikely that the field-induced Fermi surface reconstruction in $CeRhIn_5$ is caused by a change in magnetic structure. Nevertheless, it is desired to perform experiments, e.g., high-field NMR measurements, to directly examine it.

The delocalization of 4*f*-electrons, resulting from Kondo screening, expands the Fermi volume and may give rise to the emergence of larger dHvA frequencies above $B^*$ in $CeRhIn_5$. This is supported by the following facts: (1) *Evolution of dHvA frequencies*.

For $B < B^*$, previous dHvA studies have shown that the Ce-4$f$ electrons are localized in CeRhIn$_5$ and do not contribute to the Fermi sea (18, 21). Our results are compatible with this conclusion. Above $B^*(T)$, new dHvA frequencies are observed for $B//a$ and $B//c$ at low temperatures. They are comparable with both the ones measured in CeCoIn$_5$ (Ref. 22 and also SI Appendix, Fig. S6) and calculations for CeRhIn$_5$ under the assumption of itinerant 4$f$-electrons (cf. SI Appendix, Fig. S5). (2) *Change of Hall coefficient*. According to Ref. 18, the Fermi surface of LaRhIn$_5$, which has no 4$f$ electrons, is dominated by electron pockets; the Fermi surface of CeRhIn$_5$ will be similar if the 4$f$ electrons are localized, accounting for the negative Hall coefficient at fields below $B^*$. On the other hand, in CeCoIn$_5$ the 4$f$-electrons are itinerant, and the Fermi surface is more three-dimensional; the volumes of electron and hole pockets are nearly compensated (18). With delocalized 4$f$ electrons the Fermi surface of CeRhIn$_5$ would be of an analogous form, explaining the more complicated variation of the Hall coefficient above $B^*$. We note that a change in magnetic structure, e.g., a metamagnetic transition as discussed above, may lead to a jump in the Hall resistivity. However, the Hall number usually is unchanged across a metamagnetic transition, as observed at $B_M$ for CeRhIn$_5$ (see Fig. 3A). This adds further evidence against a change of magnetic structure at $B^*$. (3) *Enhancement of the heat-capacity coefficient*. In the inset of Fig. 2B, we have shown that the heat-capacity coefficient, $C_p(B)/T$, at low temperatures is substantially enhanced upon applying a magnetic field to suppress the AF order, suggesting the development of a Kondo-derived heavy electron state in a magnetic field. The persistence of the large Fermi surface up to at least $B = 70$ T at low temperatures can be understood by a simple consideration of the magnetic field and temperature scales. The temperature for the onset of Kondo screening in CeRhIn$_5$ is about 10 K (14), more than twice the Néel temperature. Given that the critical magnetic field $B_{c0}$ is about 50 T, the single-ion Kondo field is expected to be substantially larger than the presently accessible fields.

The above analyses suggest that, upon increasing the magnetic field at ambient pressure and zero temperature, (*i*) an abrupt change of Fermi surface associated with a localized-itinerant transition of the 4*f* electrons occurs at $B^* \approx 30$ T, i.e., *inside* the AF state of CeRhIn$_5$, while (*ii*) the AF QCP at $B_{c0} \approx 50$ T is presumably of the SDW-type. This is in contrast to what happens as a function of pressure at relatively low magnetic fields, where the dHvA oscillations indicate a 4*f*-localized/itinerant transition *at* the AF QCP (17). Our results are therefore two-fold. First, there are at least two different types of quantum phase transitions in CeRhIn$_5$, one induced by tuning a magnetic field and the other accessed with varying pressure. Moreover, under the tuning of these two parameters, a jump of Fermi surface occurs inside the AF state and at the AF QCP, respectively.

As a potential scenario, in Fig. 5B we present a schematic magnetic field-pressure phase diagram for CeRhIn$_5$ at zero temperature. Such a phase diagram can be qualitatively interpreted in terms of a global phase diagram of quantum-critical heavy fermion metals (23, 24). This model delineates the evolution of the zero-temperature AF transition and the Kondo-destruction in a multi-parameter phase space. While tuning the compound by a magnetic field at zero pressure, the Kondo resonances are switched on at $B_0^*[\equiv B^*(T\rightarrow 0)]$ before the AF order is suppressed at $B_{c0}$, and the Fermi surface evolves smoothly across the latter transition. This is in contrast to the transition caused by pressure at relatively low field, at which the Kondo destruction and magnetic transition take place simultaneously, leading to a jump of the Fermi surface at the zero-temperature continuous AF phase transition (17). We note that the dynamical Kondo effect makes the mass heavy even when the Fermi surface is small (3, 23). According to this model, the AF QCP associated with the pressure-induced superconducting dome is then likely to be of the unconventional type. This indicates that heavy-fermion superconductivity not only arises in the vicinity of SDW-type QCPs (7, 25), but can also be driven by electronic fluctuations arising from such an unconventional QCP at which the Kondo effect is destroyed.

Our experimental results and understandings have implications beyond $CeRhIn_5$. They can be connected to the disparate results in a variety of heavy-fermion materials. Considerable evidence already exists that pure $YbRh_2Si_2$ under magnetic-field tuning (10) and $CeCu_{6-x}Au_x$ under the variation of doping (9) feature Kondo destruction at their respective AF QCP. On the other hand, there also is evidence for Kondo destruction inside the AF region of Co-doped $YbRh_2Si_2$ (26) as well as in magnetic-field-tuned $CeCu_{6-x}Au_x$ (28) and $Ce_3Pd_{20}Si_6$ (29). In $CeIn_3$, dHvA frequencies and the corresponding cyclotron masses of heavy-hole pockets undergo a sharp increase near 40 T, which is below the critical field needed to suppress AF order to a QCP (30). This suggests the possibility of a similar localization/itinerant transition inside its AF phase, even though dHvA oscillations associated with the large Fermi surface have not yet been observed at high fields. In a related vein, it has been suggested that Kondo destruction occurs outside the AF region in materials such as Ir-doped $YbRh_2Si_2$ (26) and $Yb_2Pt_2Pb$ (27). Our Fermi-surface characterization of multiple quantum phase transitions in a single compound not only provides new understandings for $CeRhIn_5$, but also strengthens the case that the quantum phase transitions of the other electron-correlated materials may be placed on the proposed global phase diagram.

## Methods

A detailed description of the methods and materials is given in the Supporting Information. Single crystals of CeRhIn$_5$ were grown by a flux method. Room-temperature powder X-ray diffraction measurements revealed that all the crystals are single phase and crystallize in the tetragonal HoCoGa$_5$ structure. The orientation of the crystal was determined by X-ray Laue diffraction (SI Appendix, section 1). Heat capacity and dHvA oscillations were measured in pulsed field magnets at the Los Alamos National Laboratory Pulsed Field Facility (SI Appendix, section 2). A typical sample size for this study is about 1mm×0.5mm×0.1mm. An *ac* calorimeter was used for heat capacity measurements at fields up to $B = 53$ T and temperatures down to $T = 0.9$ K. Subtractions of the field-independent addenda contributions from the total heat capacity allow us to determine the absolute values of heat capacity for the sample. Magnetic susceptibility was measured by an induction method up to $B = 72$ T and at temperatures down to $T = 0.4$ K, obtained by pumping on liquid $^3$He in a $^3$He bath cryostat. The transverse Hall resistivity $\rho_{xy}(B)$ as well as the dHvA oscillations based on a torque method were measured in a $^3$He cryostat and in fields to 45T generated by the hybrid magnet at NHMFL, Tallahassee (SI Appendix, section 3-4). Band structure calculations were performed using the full-potential linearized augmented plane wave method as implemented in the WIEN2k code (SI Appendix, section 5).

**Acknowledgements.** We acknowledge valuable discussion with S. Kirchner and R. Daou. Work at Zhejiang University was supported by the National Basic Research Program of China (973 Program) (Grant Nos. 2011CBA00103 and 2009CB929104), the NSFC (Grant Nos. 11174245,10934005), the Fundamental Research Funds for the Central Universities and Zhejiang Provincial Natural Science Foundation of China. Work at LANL was performed under the auspices of the DOE and was supported by the DOE/Office of Science project "Complex Electronic Materials". Work at the National High Magnetic Field Laboratory is also supported by NSF, State of Florida and DOE BES program "Science in 100 T". TP acknowledges support from NRF (Grant No. 220-2011-1-C00014). Work at Dresden was partially supported by the DFG Research Unit 960 "Quantum Phase Transitions". Work at Rice University was in part supported by NSF (Grant No. DMR-1309531) and the Robert A. Welch Foundation (Grant No. C-1411).

**Author contributions.** H.Q.Y. designed research; L.J., Y.K., Z.W., and H.Q.Y. performed the high-field measurements with assistance from D.G., J.S., and M.J.; Samples were grown by E.D.B., T.S., H.-O.L., and J.D.T.; L.J., Y.C., E.D.B., Z.W., G.P., J.Z., and T.P. characterized the samples and partially prepared for the experiments; L.J., Y.C., Y.K., J.S., Z.W., and H.Q.Y. analyzed the experimental data; J.-X.Z. and Q.S. carried out theoretical calculations and analyses; and L.J., F.S., Q.S., and H.Q.Y. wrote the manuscript with suggestions from J.S. and J.D.T.

# References

1. Stewart GR (2001) Non-Fermi-liquid behavior in *d*- and *f*-electron metals. *Rev Mod Phys* 73:797–855.

2. von Löhneysen H, Rosch A, Vojta M, Wölfle P (2007) Fermi-liquid instabilities at magnetic quantum phase transitions. *Rev Mod Phys* 79:1015-1075.

3. Si Q, Steglich F (2010) Heavy Fermions and Quantum Phase Transitions. *Science* 329:1161-1166.

4. Hertz JA (1976) Quantum critical phenomena. *Phys Rev B* 14:1165-1184.

5. Moriya T (1985) Spin Fluctuations in Itinerant Electron Magnetism (Springer).

6. Millis AJ (1993) Effect of a nonzero temperature on quantum critical points in itinerant fermion systems. *Phys Rev B* 48:7183-7196.

7. Arndt J, et al. (2011) Spin fluctuations in normal state $CeCu_2Si_2$ on approaching the quantum critical point. *Phys Rev Lett* 106:246401.

8. Aronson MC, et al. (1995) Non-Fermi-liquid scaling of the magnetic response in $UCu_{5-x}Pd_x$ (x = 1, 1.5). *Phys Rev Lett* 75:725-728.

9. Schröder A, et al. (2000) Onset of antiferromagnetism in heavy-fermion metals. *Nature* 407:351-355.

10. Paschen S, et al. (2004) Hall-effect evolution across a heavy-fermion quantum critical point. *Nature* 432:881-885.

11. Si Q, Rabello S, Ingersent K, Smith JL (2001) Locally critical quantum phase transitions in strongly correlated metals. *Nature* 413:804-808.

12. Coleman P, Pépin C, Si Q, Ramazashvili R (2001) How do Fermi liquids get heavy and die? *J Phys.: Condens Matter* 13:R723-R738.

13. Senthil T, Vojta M, Sachdev S (2004) Weak magnetism and non-Fermi liquids near heavy-fermion critical points. *Phys Rev B* 69: 35111.

14. Hegger H, et al. (2000) Pressure-induced superconductivity in Quasi-2D $CeRhIn_5$. *Phys Rev Lett* 84: 4986-4989.

15. Park T, et al. (2006) Hidden magnetism and quantum criticality in the heavy fermion superconductor $CeRhIn_5$. *Nature* 440: 65-68.

16. Knebel G, et al. (2006) Coexistence of antiferromagnetism and superconductivity in $CeRhIn_5$ under high pressure and magnetic field. *Phys Rev B* 74:20501.

17. Shishido H, Settai R, Harima H, Ōnuki Y, (2005) A drastic change of the Fermi surface at a critical pressure in $CeRhIn_5$: dHvA study under pressure. *J Phys Soc Jpn* 74:1103-1106.

18. Shishido H, et al. (2002) Fermi Surface, Magnetic and Superconducting Properties of $LaRhIn_5$ and $CeTIn_5$ (T: Co, Rh and Ir). *J Phys Soc Jpn* 71: 162-173, and references therein.

19. Cornelius AL, et al. (2000) Anisotropic electronic and magnetic properties of the quasi-two-dimensional heavy-fermion antiferromagnet $CeRhIn_5$. *Phys Rev B* 62:14181-14185.

20. Shoenberg D (1984) *Magnetic oscillations in metals* (Cambridge University Press, Cambridge, UK).

21. Harrison N, et al. (2004) 4*f*-electron localization in $Ce_xLa_{1-x}MIn_5$ with *M*=Co, Rh, or Ir. *Phys Rev Lett* 93:186405.

22. Hall D. et al. (2001) Fermi surface of the heavy-fermion superconductor $CeCoIn_5$: The de Haas--van Alphen effect in the normal state. *Phys Rev B* 64:212508.

23. Si Q. (2006) Global magnetic phase diagram and local quantum criticality in heavy fermion

metals. *Physica B: Condensed Matter.*378-380:23-27.

24. Coleman P, Nevidomskyy AH (2010) Frustration and the Kondo effect in heavy fermion materials. *J Low Temp Phys* 161:182-202.

25. Monthoux P, Pines D, Lonzarich GG (2007) Superconductivity without phonons, *Nature* 450: 1177-1183 (2007).

26. Friedemann S, et al. (2009) Detaching the antiferromagnetic quantum critical point from the Fermi-surface reconstruction in YbRh$_2$Si$_2$. *Nature Phys* 5:465-469.

27. Kim MS, Aronson MC (2013) Spin liquids and antiferromagnetic order in the Shastry-Sutherland-lattice compound Yb$_2$Pt$_2$Pb. *Phys Rev Lett* 110: 017201.

28. Stockert O, Enderle M, von Löhneysen H (2007) Magnetic fluctuations at a field-induced quantum phase transition. *Phys Rev Lett* 99: 237203.

29. Custers J, et al. (2012) Destruction of the Kondo effect in the cubic heavy-fermion compound Ce$_3$Pd$_{20}$Si$_6$. *Nature Mat* 11:189-194.

30. Sebastian SE, et al. (2009) Heavy holes as a precursor to superconductivity in antiferromagnetic CeIn$_3$. *Proc Natl Acad Sci U.S.A.* 106:7741-7744.

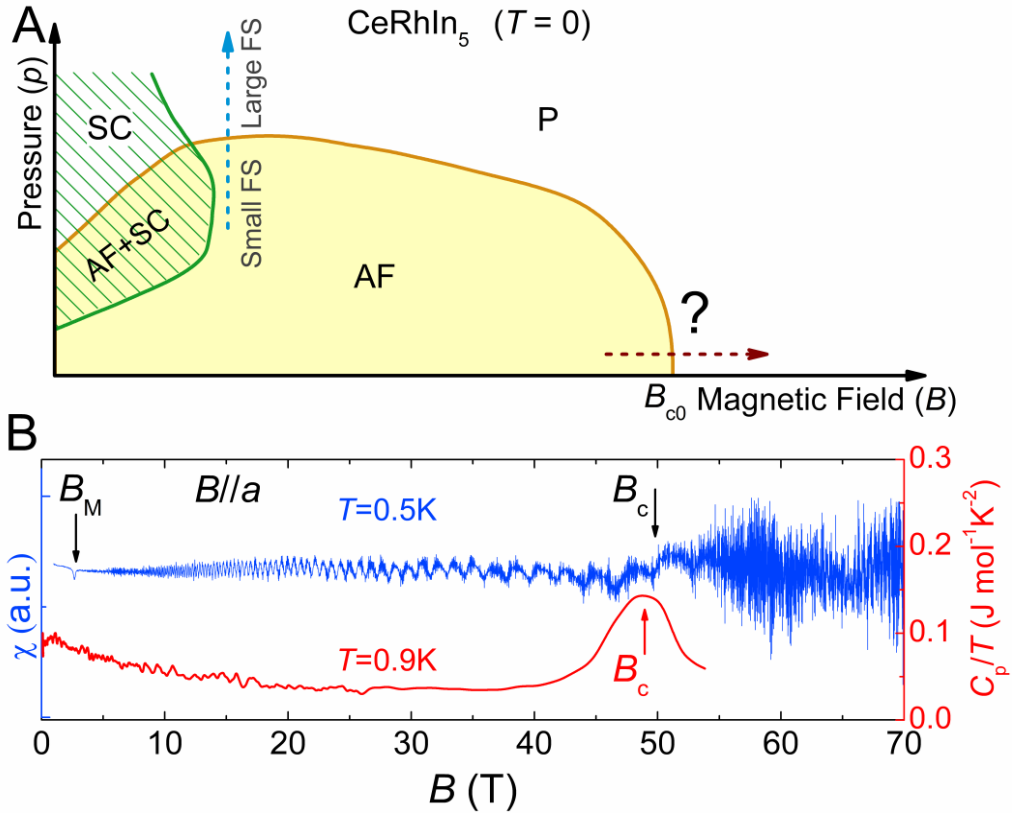

Fig. 1. Schematic phase diagram and field-dependent data for CeRhIn$_5$. (A), Schematic magnetic field-pressure phase diagram of CeRhIn$_5$ at zero temperature. Pressure suppresses the antiferromagnetic (AF) order and induces superconductivity (SC), leading to several ground states (i.e., AF order, SC and their coexistence) in the phase diagram (15, 16). A pressure-induced change from a small to a large Fermi surface and a diverging effective mass are observed at the AF QCP for fields larger than the superconducting upper critical field (17). What happens as a function of magnetic field at ambient pressure is the subject of the present study. (B), Heat capacity and dHvA oscillations in a pulsed magnetic field. The magnetic susceptibility ($T$ = 0.5 K) and the coefficient of the *ac* heat capacity $C_p/T$ ($T$ = 0.9 K) of CeRhIn$_5$ are shown as a function of magnetic field ($B$//a) up to 70 T and 53 T, respectively. They display a metamagnetic transition at $B_M \approx$ 2.5 T, and a transition from AF to paramagnetic (P) phases at a higher field $B_c(T)$. The upturn in $C_p/T$ vs. $B$ below $B$ = 15 T is due to the metamagnetic transition and also the backgrounds (see SI Appendix).

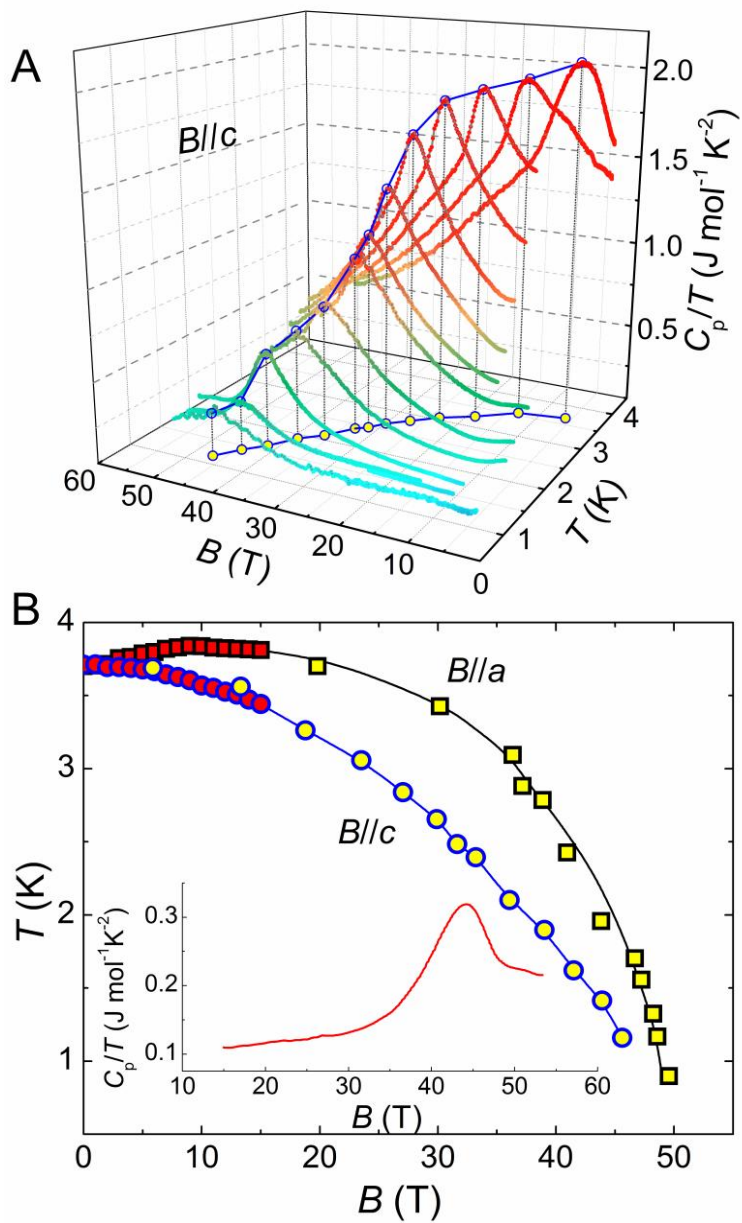

Fig. 2. Temperature and magnetic-field dependence of the heat-capacity coefficient $C_p/T$ for CeRhIn$_5$. (A), Magnetic-field dependence of $C_p/T$ at various temperatures for $B//c$. (B), Field-dependence of the Néel temperature $T_N(B)$ for $B//c$ (square) and $B//a$ (circle). Both curves extrapolate to nearly the same critical field $B_{c0}$ at $T = 0$; the red and yellow symbols represent consistent data obtained in a PPMS-16T and in pulsed fields, respectively. Inset: $C_p/T$ vs. $B$ at 1.4 K for $B//$c.

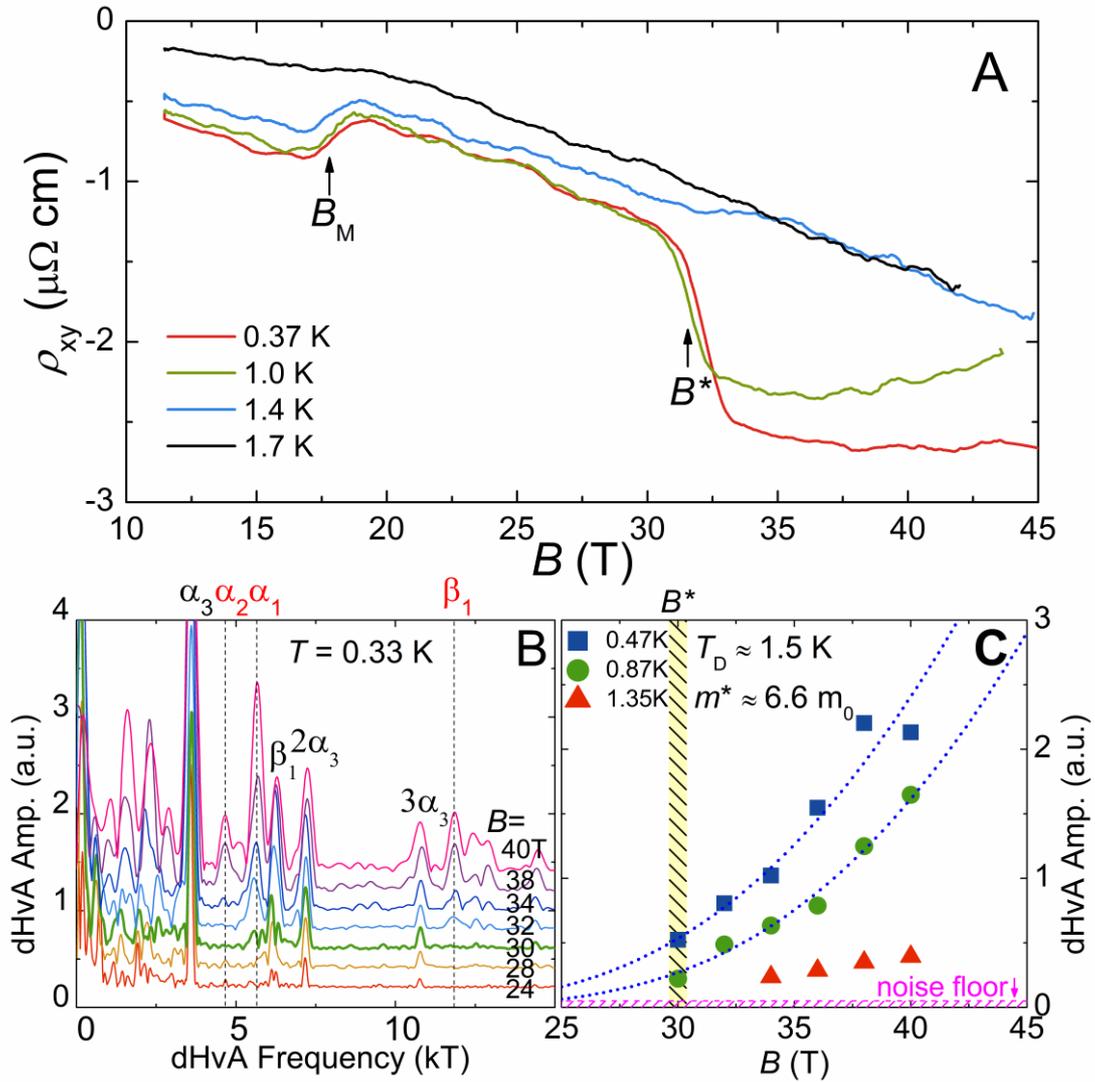

Fig. 3. Evidence of Fermi surface reconstruction for $B//c$. (A), Hall resistivity $\rho_{xy}(B)$ at different temperatures. $\rho_{xy}(B)$ shows a metamagnetic transition at $B_M \approx$ 18 T and a rapid change of the differential Hall coefficient $R_H = d\rho_{xy}/dB$ at $B = B^* \approx 31$ T; the latter suggests a sharp reconstruction of Fermi surface. Note that the metamagnetic transition at $B_M$ disappears upon rotating the sample by about 7 degrees, suggesting a small deviation of the applied magnetic field from the $c$-axis. (B), Fourier spectra of the dHvA oscillations at various magnetic fields ($T = 0.33$ K). Here a field window of $\Delta B = 5$ T is used for the Fourier analysis at each magnetic field and the Fourier spectra are shifted accordingly along the y-axis. (C), dHvA amplitude of the $\alpha_1$ branch versus magnetic field at several selected temperatures. Note that it is difficult to resolve the dHvA oscillations from the noise below the field of the first data point for each temperature. The dotted lines show fits to the LK formula (20); the fitting parameters of $T_D = 1.5$ K and $m^* = 6.6$ $m_0$ represent the Dingle temperature and the effective mass, respectively.

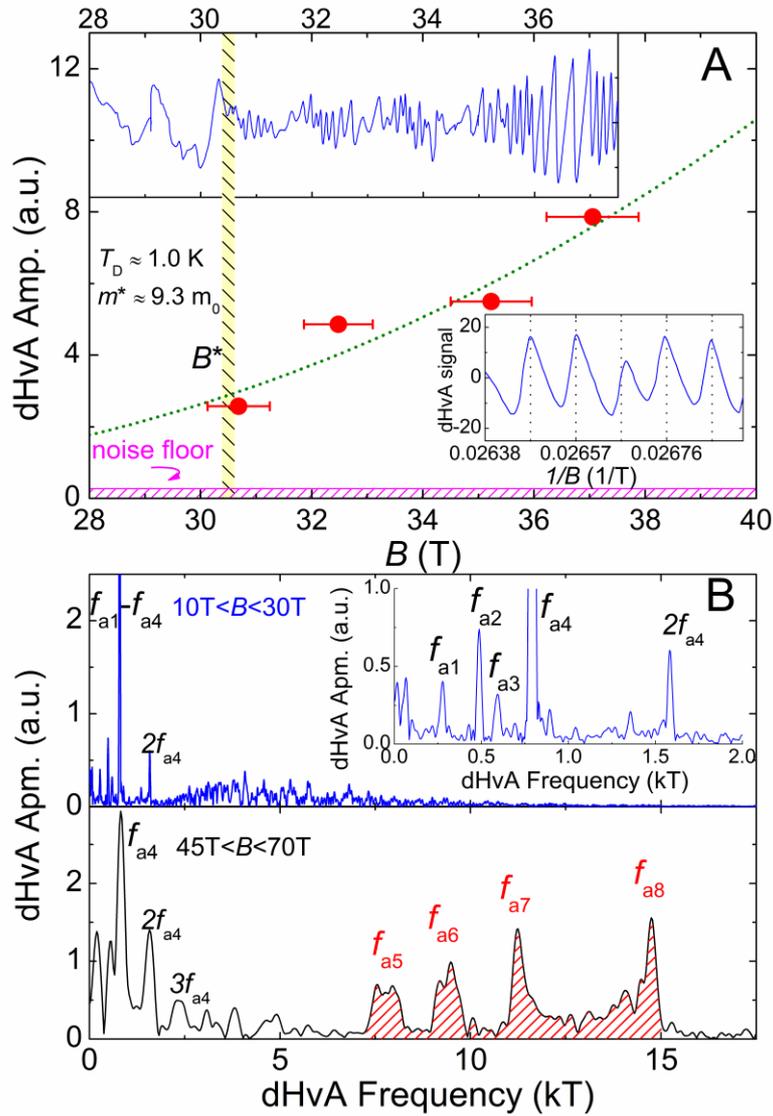

Fig. 4. Evidence of Fermi surface reconstruction for $B$//a. (A), dHvA amplitude versus magnetic field for the large dHvA frequency observed above $B^* = 30.5$ T. Data are obtained by averaging the oscillatory amplitudes over a field window of 2T centered at each presented field. Here the noise level is comparable with the symbol size. The dotted line shows a fit to the LK formula with $T_D = 1.0$ K and $m^* = 9.3$ $m_0$. The upper inset displays the dHvA oscillations after subtracting the background. The onset of a new dHvA oscillation at $B^* \approx 30.5$ T is seen clearly here as well as in the main panel. The lower inset plots periodic oscillations of dHvA signals on a $1/B$ axis. The data were measured in a 45 T hybrid magnet using a piezo-cantilever. (B), Fourier spectra of the dHvA oscillations over field windows of 10 T<$B$<30 T and 45 T<$B$<70 T, respectively. Four new dHvA frequencies are observed in the high field region. Here the data were obtained in a pulsed field up to 70 T at $T = 0.5$ K.

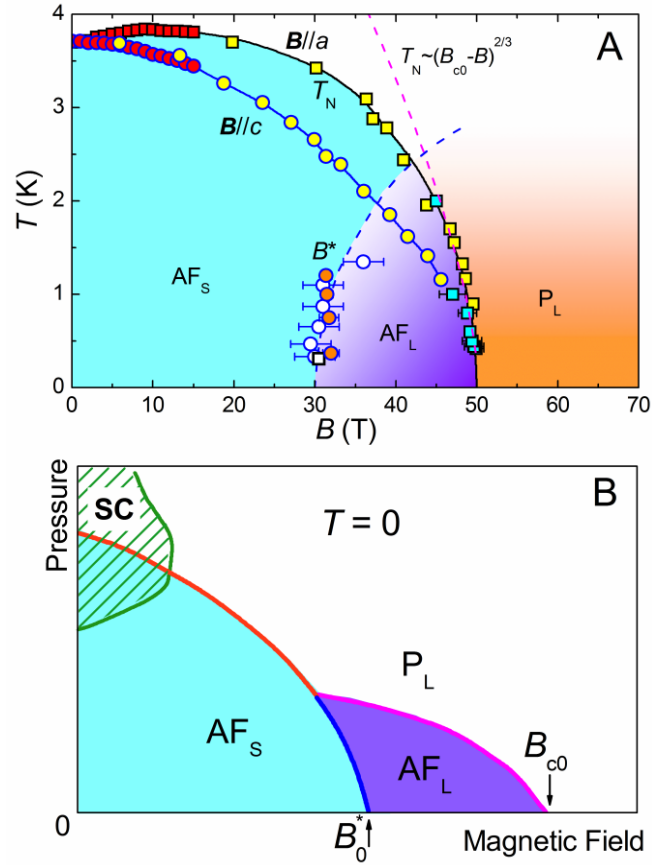

Fig. 5. Experimental temperature-field phase diagram and its relation to a multi-parameter theoretical phase diagram. (A), Temperature-magnetic field phase diagram of CeRhIn$_5$ at ambient pressure. Square and circle symbols represent the characteristic temperatures for $B//a$ and $B//c$, respectively. The Néel temperature, $T_N(B)$, was determined from the heat capacity measured in a PPMS with a dc field of 16T (red) and in a pulsed field (yellow) as well as the dHvA oscillations (light blue, $B//a$). The dashed line displays the field dependence $T_N \sim (B_{c0}-B)^{2/3}$, expected for a 3D-SDW QCP (6). The field $B^*(T)$ is determined from a jump of the Hall resistivity (orange circles, $B//c$) and the onset of new dHvA frequencies in the Fourier transform spectra (white square for $B//a$; white circles for $B//c$). The error bars were determined from the width of the Hall resistivity jumps and the field interval $\Delta B = 5$ T used in the Fourier transform analysis. The phases AF$_S$, AF$_L$ and P$_L$ represent antiferromagnetic (AF) or paramagnetic (P) states with large ("L") or small ("S") Fermi surfaces. (B), A schematic magnetic field-pressure phase diagram of CeRhIn$_5$ at $T = 0$. A pressure-induced AF QCP exists at low field and is accompanied by a sharp change of the Fermi surface, indicating a destruction of the Kondo effect at zero temperature (17). On the other hand, as shown in this work, a sufficiently strong magnetic field at ambient pressure also continuously suppresses the AF phase at $B_{c0}$. In this case, the Kondo destruction takes place inside the AF state, and the AF QCP is likely to be of a SDW-type. Further measurements are to be performed to determine experimentally the phase diagram in the regime of finite pressures and high magnetic fields.

# Supporting Information

1. **Sample preparation**

Single crystals of $CeRhIn_5$ were grown from In flux. High-purity elements Ce (99.99%), Rh (99.95%) and In (99.99%) were combined in a ratio of 1:1:20, and the mixture was placed in an alumina crucible which was sealed in an evacuated quartz ampoule. The sealed quartz ampoule was heated to 1100 °C over a period of 10 h and kept at this temperature for 4 h to ensure complete melting of all the components. The melt was cooled slowly to 700 °C at a rate of 4 °C/h. At this point, excess In flux was decanted by centrifuging. Room-temperature powder X-ray diffraction revealed that all the crystals were single phase and crystallized in the tetragonal $HoCoGa_5$ structure. The orientation of the crystal was determined by X-ray Laue diffraction.

2. **Measurements of *ac* heat capacity in a pulsed magnetic field**

The heat capacity was measured in two different pulsed magnets at Los Alamos National Laboratory: (1) a capacitor-bank driven mid-pulse magnet with a maximum field of 50 T and duration of 250 milliseconds and (2) a motor-generator-driven long-pulse magnet with a controlled waveform up to 60 T, with total pulse duration of about 2 seconds. Experimental details are described in ref. 1.

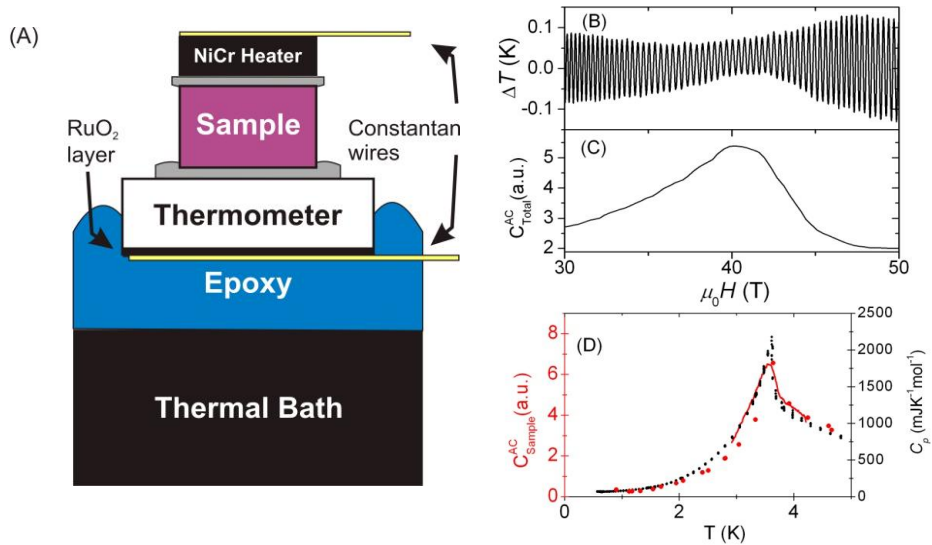

**Fig. S1** (A) Schematics for our *ac* heat capacity measurement in pulsed magnetic fields. (B) Magnetic field dependence of the temperature oscillations. (C) Total heat capacity vs. magnetic field. (D) Temperature dependence of the sample heat capacity, measured by the *ac* method in pulsed fields (red symbols) and by the PPMS (black symbols) at zero field. The data shown in (B) and (C) were measured at $T = 2.1$K for $B//a$.

Fig. S1(A) shows a schematic drawing of the *ac* calorimeter used for these studies. A single crystalline sample of CeRhIn$_5$ was mounted on the back of a RuO$_2$ thermometer (sapphire substrate) with small amount of silver paint. A NiCr thin film was deposited on a small piece of silicon single crystal, and mounted on the top of the sample. The RuO$_2$ thermometer was thermally and electrically isolated from a silicon single crystal holder (and thermal bath) by a thin layer of epoxy resin (Stycast 2850). For temperature regulation, the thermal link between the sample and the thermal bath was tuned by adding a small amount of $^4$He or $^3$He gas whose contribution to the total heat capacity was negligible. In this configuration we measure the total heat capacity ($C_{\text{total}}$) as the sum of the sample ($C_{\text{sample}}$) and addenda heat capacities ($C_{\text{addenda}}$). Here, the contributions to the addenda are due to the heater, thermometer and epoxy layer. When *ac* power ($P_{\text{ac}}$) is applied to the heater, the sample temperature ($T_{\text{ac}}$) oscillates at a typical frequency of $f = 1$ kHz. Fig. S1(B) shows an example of the temperature oscillations measured at $T = 2.1$ K for $B//a$. By numerically analyzing the $T_{\text{ac}}$ data, we extract the amplitude of $|T_{\text{ac}}|$ and the phase delay, $\varphi$, between $T_{\text{ac}}$ and $P_{\text{ac}}$. Then the total heat capacity $C_{\text{total}}$ is calculated as $C_{\text{total}} = P_{\text{ac}} \sin\varphi / |T_{\text{ac}}| 2\pi f$ [1], the field dependence of which is plotted in Fig. S1(C). By assuming that the addenda contribution is given by the expression $C_{\text{addenda}} = \alpha T + \beta T^3 + \gamma T^5$, the sample heat capacity is extracted by fitting and normalizing $C_{\text{sample}}$ (= $C_{\text{total}}$ - $C_{\text{addenda}}$) to the heat capacity $C_p(T)$ obtained using a relaxation method in a commercial Quantum Design® Physical Properties Measurement System (PPMS). The so-derived sample heat capacity $C_{\text{sample}}$ at zero field is plotted as a function of temperature in Fig. S1(D). Assuming that the addenda contributions are independent of magnetic field, we estimate the absolute value of the heat capacity of the sample in pulsed fields following the above procedures. The heat capacity divided by temperature $C_p(T)/T$ obtained in pulsed magnetic fields is remarkably close to that measured in *dc* fields by Kim et. al. [2], and could be scaled nicely by the Néel temperature $T_N$ as shown in Fig. S2, indicating the validity of our method. The shape of the $C_p/T(B)$ anomaly evolves from clearly asymmetric at $T = 3.7$ K to somewhat more symmetric at temperatures $T < 2$ K. The pronounced field-asymmetry observed at high temperatures is likely due to the contributions of classical fluctuations inside the ordered state $B < B_c(T)$. The asymmetry at lower temperatures, on the other hand, is expected and characteristic for a second order phase transition. Furthermore, we do not see added accumulation of entropy around $T_N(B)$, a hallmark of first order phase transitions [3]. Hence, a change of order in the nature of the phase transition as the temperature is reduced can be excluded in this case.

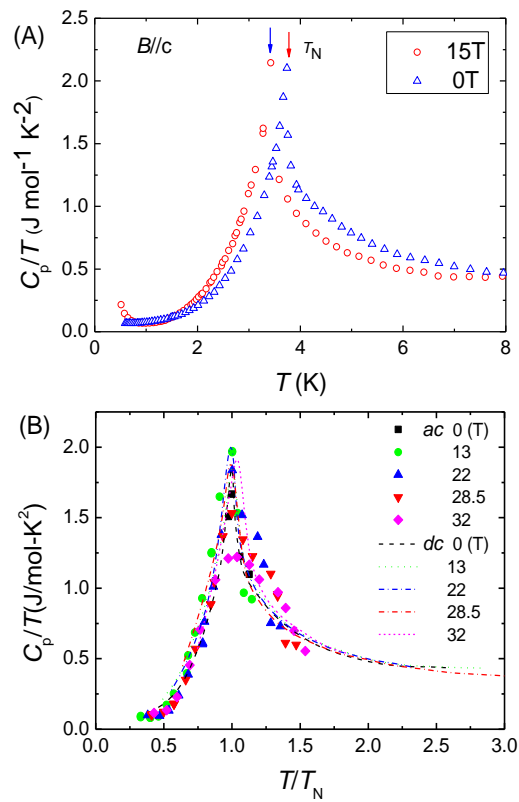

**Fig. S2** Temperature dependence of the heat-capacity coefficient $C_p/T$ for CeRhIn$_5$ ($B//c$): (A) at $B = 0$ and 15 T measured in a Quantum Design® PPMS; (B) at various fields obtained in a *dc* magnet (data from ref. 2) and in a pulsed magnetic fields (symbols; this study). In the latter case, the temperatures are normalized by the corresponding field-dependent Néel temperature. $T_N$ is defined as the temperature of the maximum in $C_p/T$ *vs* $T$. The lambda-type anomaly indicates a second-order phase transition.

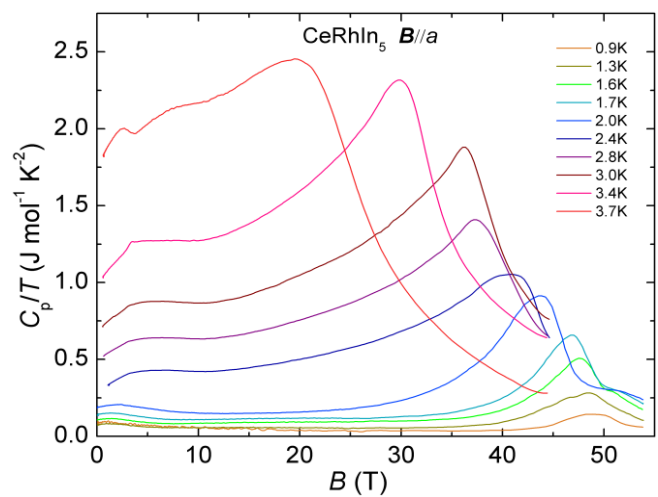

**Fig. S3** Magnetic field dependence of the heat capacity divided by temperature for CeRhIn$_5$ at various temperatures for $B//a$.

In Fig. S3, we show the magnetic field dependence of $C_p/T$ for CeRhIn$_5$ at various temperatures and for the magnetic field $B//a$-axis. Low field anomalies around 2.5 T indicate a metamagnetic transition as described in the main text. The pronounced maximum in $C_p/T(B)$ marks the onset of the antiferromagnetic transition which shifts to higher field with decreasing temperature. We note that $T_N(B)$, determined from the peaks of $C_p(T)/T$ measured at fixed magnetic fields in the Quantum Design® PPMS-16T and in a quasistatic 35T magnet (cf. Fig. S2), is highly consistent with our heat capacity data obtained in pulsed magnetic fields at constant temperatures (see the main text).

3. **Measurements of Hall resistivity**

The Hall resistivity was measured by using a standard four-point method in the 45T hybrid magnet at NHMFL Tallahassee FL, USA. The electrical contacts were prepared by using the spot welding method. In this study, a magnetic field was applied along the $c$-axis and a longitude current of 10mA flows in the $ab$-plane of a thin crystal with dimensions of 2mm×2mm×0.2mm. The transverse Hall voltage was measured by using a LakeShore 370 ac resistance bridge. During the measurements, the sample was immersed in $^3$He liquid. Great care was taken to ensure that there was no sample heating by checking that the result was independent of the excitation current. The voltage contacts are well aligned perpendicularly with respect to current. Thus, the contributions from the magnetoresistance to the Hall signals are negligible, which was carefully examined by rotating the sample by 180°.

4. **de Haas-van Alphen (dHvA) effect: measurements and data analyses**

As a consequence of the quantization of closed electronic orbits in a magnetic field, the magnetic susceptibility in a metal periodically oscillates as a function of the reciprocal magnetic field ($1/B$). The oscillatory frequencies are proportional to the extremal areas of the Fermi surface, i.e., $f_i = 1/\Delta(1/B) = \hbar A_i/2\pi e$. Here $f_i$ is the dHvA frequency, $e$ is the charge of the electron and $A_i$ the extremal cross-section of the $i$th branch of the Fermi surface in a plane perpendicular to the magnetic field.

In this study, we measured the dHvA oscillations using two different techniques, i.e., a torque method in a quasi-static magnet up to 45 T and an induction method in a pulsed magnetic field up to 72 T. Fig. S4 presents the dHvA signals obtained in a pulsed field at $T = 0.42$K, 0.5K and 0.6K for $B//a$, which show clear quantum oscillations. As seen in Fig. S4, there is a metamagnetic transition at $B_M$ and an AF transition at $B_c$ as marked by the arrows. Error bars on $B_c$(T) shown in Fig. 5A (main text) are the result of uncertainty in determining its value from the construction illustrated in Fig. S4.

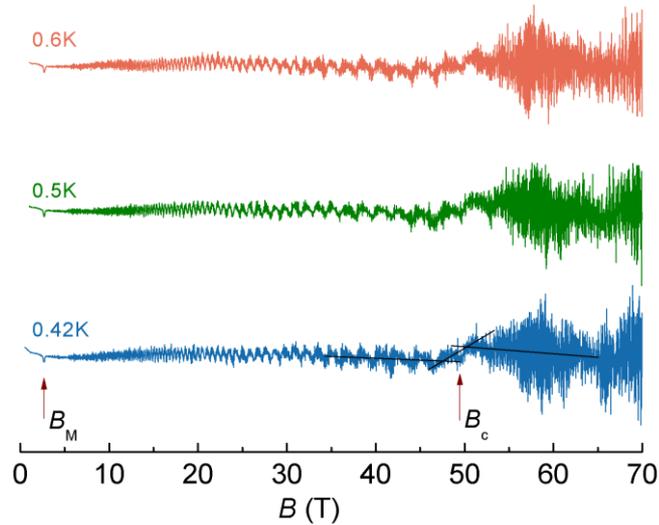

**Fig. S4** Magnetic susceptibility plotted as a function of magnetic field at various temperatures. $B_M$ and $B_c$ mark the critical field for the metamagnetic transition and the AF transition, respectively.

The dHvA frequencies $f_i(B)$ were determined by performing a numerical Fourier transform of the dHvA oscillations over a field interval of $[B-\Delta B/2, B+\Delta B/2]$. In order to show the evolution of the dHvA frequencies as a function of magnetic field, we carefully analyzed the Fourier spectra obtained at various magnetic fields and over various magnetic field windows of width $\Delta B$. Narrowing the magnetic field windows degrades the resolution of the Fourier spectra.

## 5. Band structure calculations of CeRhIn$_5$

In order to understand the dHvA frequencies observed in the different magnetic-field regimes, we calculated the dHvA frequencies using the Fermi surfaces of CeRhIn$_5$ determined from first-principles band structure calculations. Such calculations were carried out assuming the Ce 4$f$ electrons to be either localized or itinerant. Initial band structure calculations for CeRhIn$_5$ were performed by taking the 4$f$ electrons as itinerant [4], and the dHvA frequencies were obtained for magnetic field oriented along both [100] and [001] directions. A localized nature of the 4$f$ electrons in CeRhIn$_5$, however, was proposed based on a comparison of its dHvA spectrum to that of LaRhIn$_5$, which has no 4$f$ electron [5]. This was further supported by direct calculations of the dHvA frequencies that assumed the Ce 4$f$-electrons to be localized in CeRhIn$_5$, which provided a better description of the experimental data [6]. Considering that, in the present experiment, we applied the magnetic field along the [100] direction and that no corresponding dHvA frequencies have been calculated from the band structure for the case of localized Ce 4$f$ scenario, we performed a systematic study of the electronic structure for CeRhIn$_5$ within the framework of density functional theory. The full-potential linearized augmented plane wave method as implemented in the WIEN2k code [7] was used. The spin-orbit coupling was included in a second-variational procedure. A generalized gradient approximation of Perdew, Burke, and Ernzerhof [8] was used for the treatment of exchange and correlation interactions. The energy threshold to separate the localized and delocalized electronic states was chosen to be -6 Ryd. The muffin-tin radii are 2.5 $a_0$ for Ce, 2.5 $a_0$ for Rh, and 2.43 $a_0$ for In, where $a_0$ is the Bohr radius. The extremal Fermi-surface cross sections were calculated using the Supercell K-space Extremal Area Finder, Version 1.2.0.r124 [9]. We should note that such an electronic structure calculation with core-4f electrons does not capture the dynamical competition between the RKKY and Kondo interactions from which an antiferromagnetic phase with a small Fermi surface (AF$_S$) has been theoretically determined in Kondo lattice models. However, because Fermi surface is a ground-state property, the core-4f-based electronic structure calculation is expected to well describe the Fermi surfaces of the AF$_S$ phase, under the standard assumption that the electron self-energy has only a smooth dependence on momentum. We also note that the electronic structure calculations are done in the paramagnetic state, so its comparison with the experimentally measured dHvA frequencies is appropriate for sheets of the Fermi surfaces that either do not touch the AF-zone boundary, or acquire only small energy gaps which are then magnetically broken down in the field range.

**Fig. S5** Theoretical Fermi surfaces for CeRhIn$_5$, where the Ce 4$f$ electrons are treated as (a) localized and (b) itinerant.

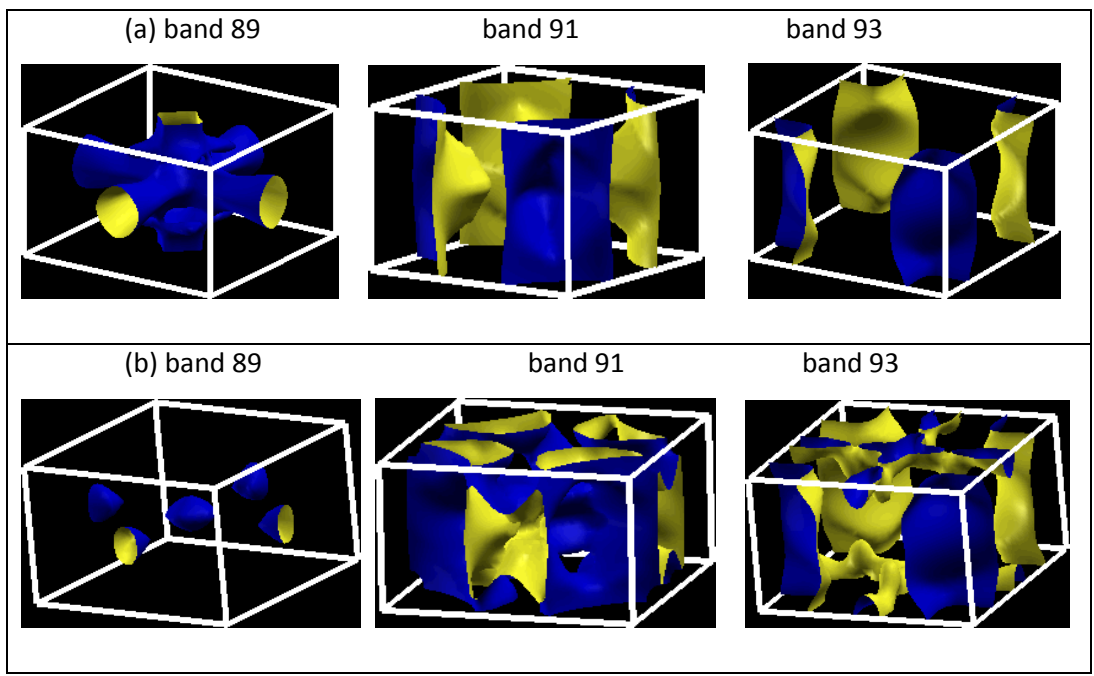

In Fig. S5, we present the calculated Fermi surfaces of CeRhIn$_5$ with Ce 4$f$-electrons treated as localized (a) and itinerant (b), respectively. In both cases, bands 89, 91 and 93 contribute to the Fermi surfaces. Results for CeRhIn$_5$ in the localized 4$f$ case are similar to those for LaRhIn$_5$, while the results for CeRhIn$_5$ in the itinerant 4$f$ case are similar to those for CeCoIn$_5$ [5]. This observation is in good agreement with the fact that there are no 4$f$ electrons to contribute to the Fermi surface of LaRhIn$_5$, while itinerant 4$f$ electrons do so in CeCoIn$_5$. As pointed out in ref. 5, the hole Fermi surface (band 89) in the case of localized 4$f$-electrons in CeRhIn$_5$ shrinks and changes into two small closed Fermi surfaces when the single 4$f$ electron in each Ce site is delocalized and becomes a conduction electron. Simultaneously, it was found that the Fermi surface topology in bands 91 and 93 evolves from a two-dimensional character in 4$f$-localized CeRhIn$_5$ to a more complicated 3D-structure in 4$f$-itinerant CeRhIn$_5$. This observation has a profound implication on the dHvA frequencies when the magnetic field is applied along the *a*-axis.

**Tab. S1** Calculated dHvA frequencies for CeRhIn$_5$ with Ce 4$f$-electrons being treated as localized (4$f$-loc.) and itinerant (4$f$-it.). Results for field along [100] and [001] as well as for a deviation of 8 degrees from [100] are shown.

| Band | $f$(T) ($B//a$) (4$f$-loc.) | $f$(T) (8° from $B//a$) (4$f$–loc.) | $f$(T) ($B//c$) (4$f$-loc.) | $f$(T) ($B//a$) (4$f$-it.) | $f$(T) ($B//c$) (4$f$–it.) | $f$(T) (8° from $B//a$) (4$f$–it.) | $m$ (m$_0$) (8° from $B//a$) (4$f$–it.) |
|---|---|---|---|---|---|---|---|
| 89 | 370.3 | 371.0 | 589.8 | 335.2 | 487.8 | 335.6 | 0.7274 |
|  | 448.4 | 471.7 | 2119.9 | 475.3 | 782.0 | 481.7 | 0.5077 |
|  | 814.4 | 818.7 | 12168.4 | 483.3 |  | 487.6 | 1.4538 |
|  | 1175.6 | 895.1 |  | 500.9 |  | 495.7 | 1.4117 |
|  | 5490.3 | 1222.2 |  |  |  |  |  |
| 91 |  |  | 6196.9 | 188.1 | 886.8 | 527.9 | 3.1104 |
|  |  |  | 6359.7 | 481.0 | 1282.4 | 1398.9 | 1.4073 |
|  |  |  | 10208.2 | 1362.6 | 1459.0 | 1580.8 | 2.5763 |
|  |  |  |  | 1554.3 | 7050.9 | 1613.1 | 1.9048 |
|  |  |  |  | 1629.8 | 12155.7 | 2344.7 | 1.9633 |
|  |  |  |  | 2274.6 | 12492.5 | 3546.6 | 4.0524 |
|  |  |  |  |  |  | 6822.4 | 8.1756 |
|  |  |  |  |  |  | 9362.8 | 5.4283 |
|  |  |  |  |  |  | 11149.4 | 5.872 |
|  |  |  |  |  |  | 11528.8 | 7.6298 |
|  |  |  |  |  |  | 13226.8 | 6.5922 |
| 93 |  |  | 3459.3 | 160.0 | 1093.7 | 179.8 | 0.6055 |
|  |  |  | 4040.4 | 175.2 | 1257.6 | 202.3 | 0.8729 |
|  |  |  | 4772.2 | 249.7 | 3808.9 | 262.2 | 0.7871 |
|  |  | 30087.4 |  | 304.2 | 4300.2 | 318.8 | 1.7382 |
|  |  |  |  | 384.6 | 5508.9 | 386.3 | 0.4372 |
|  |  |  |  | 459.7 |  | 436.5 | 1.2577 |
|  |  |  |  |  |  | 3286.1 | 6.1783 |

In Tab. S1, we compare the calculated dHvA frequencies for Ce 4*f* electrons being localized and itinerant at several magnetic field orientations. The dHvA frequencies, obtained by assuming itinerant 4*f* electrons, are in reasonable agreement with ref. 4 for the magnetic field along both [100] and [001] directions; whereas, those obtained by assuming localized 4*f* electrons are comparable to results in ref. 6, where the dHvA frequencies were only calculated for *B*//[001]. The reasonable agreement with previous calculations for comparable conditions serves as a consistency check of the present calculations. One very important result from our calculations is that, for the magnetic field applied along [100], dHvA frequencies are associated solely with band 89 and not at all with the bands 91 and 93 when Ce 4*f* electrons are localized. On the other hand, for the case of itinerant Ce 4*f* electrons, all these bands contribute to the dHvA frequencies. Our theoretical results are consistent with the experimental observation of an increase in the number of frequencies when the magnetic field exceeds $B^*$, suggesting a field-driven transition of Ce 4*f* electrons from localized to itinerant states in CeRhIn$_5$. The theoretical dHvA values for CeRhIn$_5$ with localized Ce 4*f* electrons are somewhat larger than the experimentally measured ones at low fields ($B < B^*$) (see Tab. S2), which could be due to the fact that CeRhIn$_5$ is in an antiferromagnetic state but calculations are made for a paramagnetic state. We note that the dHvA frequencies can be sensitive to field misalignment when the field is applied along the [100] direction in the case of itinerant Ce 4*f* electrons because of the complicated Fermi surface with nearly open orbits along [001] (see Fig. S5). With this in mind, we also performed calculations of dHvA frequencies at several magnetic field orientations slightly away from [100]. In Tab. S1, we list the dHvA frequencies and their corresponding effective masses calculated for a magnetic field applied 8 degrees away from the [100]

**Tab. S2:** dHvA frequencies of CeRhIn$_5$ in the low-field region (*B*//*a*, pulsed field data at $T = 0.5$ K). Experimental values are determined by performing a fast Fourier transform (FFT) over the magnetic field of 10 T-40 T ($B < B^*$). The theoretical dHvA frequencies and the corresponding effective masses are calculated by treating Ce 4*f*-electrons as localized. The magnetic field is assumed to deviate by 8 degrees from the [100] direction toward the [001] direction in these calculations (the same for Tab. S3).

| Branch | $f$(T) (*exp.*) | $f$(T) (*cal.*) | $m$ ($m_0$) (*cal.*) |
|---|---|---|---|
| $f_{a1}$ | 275 | 371 | 0.290 |
| $f_{a2}$ | 487 | 472 | 0.200 |
| $f_{a3}$ | 597 | 819 | 0.213 |
| $f_{a4}$ | 792 | 895 | 0.729 |
| | | 1222 | 0.736 |

28**Tab. S3** The large dHvA frequencies of CeRhIn$_5$ in the high field region ($B//a$, pulsed field data at $T = 0.5$ K). The experimental values are determined by performing a FFT over a magnetic field of 50 T-70 T ($B > B^*$). The theoretical dHvA frequencies and the corresponding effective masses are calculated by treating Ce 4$f$-electrons as itinerant. Here only the large dHvA frequencies occurring at $B > B^*$ are shown for comparison.

| Branch | $f$(T) (*exp.*) | $f$(T) (*cal.*) | $m$ ($m_0$) (*cal.*) |
|---|---|---|---|
| $f_{a5}$ | 7756 | 6822 | 8.176 |
| $f_{a6}$ | 9589 | 9363 | 5.428 |
| $f_{a7}$ | 11274 | 11150 | 5.872 |
| $f_{a8}$ | 14666 | 13227 | 6.592 |

direction. Such a field misorientation is quite possible in our experiments. In the itinerant case, large dHvA frequencies of 7 kT - 15 kT are reasonably close to the new dHvA frequencies experimentally observed at $B > B^*$ (see Tab. S3). Note that these large frequencies are absent if we assume the Ce 4$f$ electrons to be localized. In summary, for the field close to the $ab$-plane, we have shown that the distinction between the Fermi surfaces of the itinerant and localized 4$f$-electron cases is reflected in a particularly prominent way. Here, the dHvA frequencies for the 4$f$-localized Fermi surfaces are confined to a relatively small frequency range, but those for 4$f$-itinerant Fermi surfaces contain branches at larger frequencies. This is consistent with the change of the dHvA frequencies as the magnetic field is increased through $B^*$. Although quantitative comparisons of calculated and measured dHvA frequencies are always difficult, our calculations give quite reasonable agreement with experiment. In addition, it is a known fact that the effective masses obtained from the density functional theory with local density of generalized gradient approximation are underestimated due to the inadequacy of these approximations to capture electronic correlations. However, according to the Fermi liquid picture, the Fermi surface geometry should still be captured reasonably well by the density functional theory. Taken together, these results provide compelling evidence that the observed Fermi-surface reconstruction originates from a field-induced localization-delocalization transition of the Ce 4$f$ electrons in CeRhIn$_5$.

### 6. Comparison of the large Fermi surfaces of CeRhIn$_5$ and CeCoIn$_5$

To further address the nature of the additional dHvA frequencies of CeRhIn$_5$ at $B > B^*$, we compare them with those measured in CeCoIn$_5$ in Fig. S6. The main figure replots the dHvA frequencies of CeCoIn$_5$ from ref. 10. The major dHvA frequencies of CeRhIn$_5$ observed at $B > B^*$ for $B//a$ and $B//c$ are denoted by the filled symbols. As stated in the preceding section, the large dHvA frequencies are extremely sensitive to the field misalignment for $B//a$. As a result, we slightly shifted the angle away from $B//a$ in order to make the four new dHvA frequencies compatible with our band structure calculations of the itinerant case in the preceding section. One can see that the dHvA frequencies of CeRhIn$_5$ at $B > B^*$, either for $B//a$ or $B//c$, are reasonably compatible with those of CeCoIn$_5$. This is in contrast with the low field regime where dHvA frequencies compare well with those measured in LaRhIn$_5$ [5]. It has been well established that the Fermi surfaces of these two reference compounds (CeCoIn$_5$ and LaRhIn$_5$) are respectively large and small; in the former case the itinerant Ce 4$f$ electrons contribute to the Fermi sea [5]. Such a consistency provides additional evidence that the change of Fermi surface at $B = B^*$ is attributed to a field-induced delocalization of Ce 4$f$ electrons in CeRhIn$_5$.

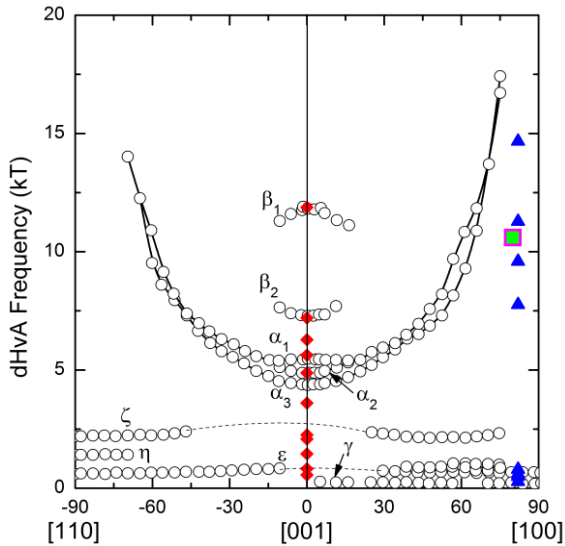

**Fig. S6** Comparison of the dHvA frequencies of CeRhIn$_5$ (at $B > B^*$) and CeCoIn$_5$. The filled and empty symbols are for CeRhIn$_5$ and CeCoIn$_5$, respectively. The filled green square represents the data from a dc field for $B$//a. Data for

7. **The schematic magnetic field-pressure phase diagram of CeRhIn$_5$ and its relation to a multi-parameter theoretical phase diagram.**

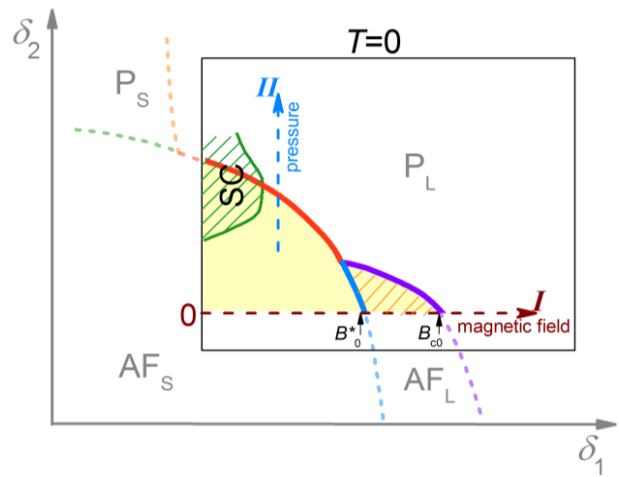

**Fig. S7** Placing CeRhIn$_5$ in the theoretically proposed global phase diagram for heavy-fermion metals. The outer layer with faded color shows the theoretical global phase diagram [11]. Embedded in it is a schematic magnetic field-pressure phase diagram of CeRhIn$_5$ at $T = 0$. The phases AF$_S$, AF$_L$, P$_S$ and P$_L$ represent antiferromagnetic (AF) or paramagnetic (P) states with large ("L") or small ("S") Fermi surfaces. A pressure-induced AF QCP exists at low field and is accompanied by a sharp change of the Fermi

surface, indicating a destruction of the Kondo effect at zero temperature [12]. On the other hand, as shown in this work, a sufficiently strong magnetic field at ambient pressure also continuously suppresses the AF phase at $B_{c0}$. In this case, the Kondo destruction takes place inside the AF state (at $B_0^*$), and the AF QCP is likely to be of an SDW-type.

**References:**

1. Kohama, Y., Marcenat, C., Klein, T. & Jaime, M. AC measurement of heat capacity and magnetocaloric effect for pulsed magnetic fields. *Rev Sci Instrum* **81**, 104902 (2010).

2. Kim, J.S., Alwood, J., Stewart, G.R., Sarrao, J.L. & Thompson, J.D. Specific heat in high magnetic fields and non-Fermi-liquid behavior in CeMIn$_5$ (M=Ir, Co). *Phys. Rev. B* **64**, 134524 (2001).

3. Jaime, M., Kim, K.H., Jorge, G., McCall, S. & Mydosh, J.A. High Magnetic Field Studies of the Hidden Order Transition in URu$_2$Si$_2$. *Phys. Rev. Lett.* **89**, 287201 (2002).

4. Hall, D., *et al.*, Electronic structure of CeRhIn$_5$: de Haas-van Alphen and energy band calculations. *Phys. Rev. B* **64**, 64506 (2001).

5. Shishido, H., *et al.*, Fermi Surface, Magnetic and Superconducting Properties of LaRhIn$_5$ and CeTIn$_5$ (T: Co, Rh and Ir). *J Phys Soc Jpn* **71**, 162 (2002).

6. Elgazzar, S., Opahle, I., Hayn, R. & Oppeneer, P.M. Calculated de Haas-van Alphen quantities of CeMIn$_5$ (M=Co, Rh, and Ir) compounds. *Phys. Rev. B* **69**, 214510 (2004).

7. Blaha, P., *et al.*, *An Augmented Plane Wave + Local Orbitals Program for Calculating Crystal Properties* (Schwarz K., Tech. Universitat Wien, Austria, 2001).

8. Perdew, J.P., Burke, K. & Ernzerhof, M. Generalized Gradient Approximation Made Simple. *Phys. Rev. Lett.* **77**, 3865 (1996).

9. Rourke, P.M.C. & Julian, S.R. Numerical extraction of de Haas–van Alphen frequencies from calculated band energies. *Comput. Phys. Commun.* **183**, 324 (2012).

10. Hall, D., *et al.*. Fermi surface of the heavy-fermion superconductor CeCoIn$_5$: The de Haas-van Alphen effect in the normal state. *Phys. Rev. B* **64**, 212508 (2001).

11. Si, Q. Quantum criticality and global phase diagram of magnetic heavy fermions. *Phys. Status Solidi. B* **247**, 476 (2010).

12. Shishido H, Settai R, Harima H & Ōnuki Y. A drastic change of the Fermi surface at a critical pressure in CeRhIn$_5$: dHvA study under pressure. *J. Phys. Soc. Jpn.* **74**, 1103 (2005).